\documentclass[draft]{agujournal2019}
\usepackage{url} 
\usepackage{lineno}
\usepackage[inline]{trackchanges} 
\usepackage{soul}
\usepackage{graphicx}%
\usepackage{amsmath,amssymb,amsfonts}%
\usepackage{amsthm}%
\usepackage{mathrsfs}%
\usepackage[title]{appendix}%
\usepackage{xcolor}%
\usepackage{textcomp}%
\usepackage{manyfoot}%
\usepackage{ragged2e}%
\usepackage{booktabs,makecell,multirow,tabularx}%
\usepackage{algorithm}%
\usepackage{algorithmicx}%
\usepackage{algpseudocode}%
\usepackage{listings}%
\usepackage{subfig}
\usepackage{epstopdf}
\usepackage{graphicx}
\usepackage[strings]{underscore}
\usepackage{setspace}
\draftfalse

\journalname{JGR: Oceans}

\begin{document}

%
%


\title{STNet: Prediction of Underwater Sound Speed Profiles with An Advanced Semi-Transformer Neural Network}
%
%




\authors{Wei Huang\affil{1}, Jiajun Lu\affil{1}, Hao Zhang\affil{1}*, Tianhe Xu\affil{2}*}

\affiliation{1}{Faculty of Information Science and Engineering, Ocean University of China, Songling Road, 266100, Qingdao Shandong, China}

\affiliation{2}{School of Space Science and Technology, Shandong University (Weihai), Wenhua West Road, 264200, Weihai, Shandong, China}




\correspondingauthor{Hao Zhang; Tianhe Xu}{zhanghao@ouc.edu.cn; thxu@sdu.edu.cn}



\begin{keypoints}
\item A semi-transformer neural network model was proposed for accurate and real-time prediction of ocean sound speed profiles.
\item A parallel processing strategy based on time and position encoding was proposed to accelerate training process.
\item An ocean experiment in the South China Sea in April 2023 was conducted for evaluate the effectiveness of the model.
\end{keypoints}

%
%

%
%


\begin{abstract}
Real time acquisition of accurate underwater sound velocity profile (SSP) is crucial for tracking the propagation trajectory of underwater acoustic signals, making it play a key role in ocean communication positioning. SSPs can be directly measured by instruments or inverted leveraging sound field data. Although measurement techniques provide a good accuracy, they are constrained by limited spatial coverage and require substantial time investment. The inversion method based on real-time measurement of acoustic field data improves operational efficiency, but loses the accuracy of SSP estimation and suffers from limited spatial applicability due to its stringent requirements for ocean observation infrastructure. To achieve accurate long-term ocean SSP estimation independent of real-time underwater data measurements, we propose a Semi-Transformer neural network (STNet) specifically designed for simulating sound velocity distribution patterns from the perspective of time series prediction. The proposed network architecture incorporates an optimized self-attention mechanism to effectively capture long-range temporal dependencies within historical sound velocity time-series data, facilitating accurate estimation of current SSPs or prediction of future SSPs. Through architectural optimization of the Transformer framework and integration of a time encoding mechanism, STNet could effectively improve computational efficiency. Comparative experimental results reveal that STNet outperforms state-of-the-art models in predictive accuracy and maintain good computational efficiency, demonstrating its potential for enabling accurate long-term full-depth ocean SSP forecasting.
\end{abstract}

 \section{Introduction}
The integrated underwater positioning, navigation, timing and communication (PNTC) system is of great significance in marine disaster warning, rescue operations and resource exploration \cite{Huang2024SurveySSP}. Acoustic signals exhibit significantly lower energy attenuation in underwater environment than radio waves, with propagation capabilities extending to tens of kilometers, thereby establishing themselves as the optimal signal carriers for underwater PNTC systems \cite{Erol2011SurveyLoc,Luo2021LocReview}. However, the heterogeneous nature of underwater sound velocity distribution induces signal propagation path curvature in accordance with Snell's law when traversing different depth layers, consequently compromising distance measurement accuracy and ultimately degrading the performance of acoustic positioning systems. Therefore, the rapid acquisition of regional sound velocity distribution becomes very important that it could help reconstructing real-time signal propagation path through ray-tracing theory, which enhances ranging and positioning precision by calculating the equivalent line-of-sight distances \cite{Liu2024Loc,Huang2024Loc}.

The velocity of underwater acoustic signal propagation is determined by multiple environmental parameters, primarily temperature, salinity, and static pressure \cite{Munk1983Tomography}. In shallow marine environments, seasonal and diurnal variations induce substantial temperature fluctuations, causing significant spatiotemporal nonlinearity in sound velocity distribution. Conversely, in deep ocean regions, static pressure dominates as the primary factor influencing sound velocity, exhibiting an approximately linear relationship with increasing depth. In general, the vertical gradient variation of sound velocity is more obvious than the horizontal variation, making the sound speed profile (SSP) a standard for characterizing the distribution of underwater acoustic velocity \cite{Jensen2011COA}. In other words, an SSP actually represents a vertical distribution of acoustic propagation velocities at discrete depth intervals within a specific geographic region \cite{Liu2020UTarTrack,Huang2021SSPInversion}.

The acquisition of underwater SSPs primarily employs two methodological approaches: direct measurement techniques and inversion-based estimation methods \cite{Huang2024SurveySSP}. The direct measurement method mainly relies on instruments such as sound velocity profiler (SVP) \cite{Zhang2022SVP} or conductivity, temperature, and depth profiler (CTD) \cite{Luo2023CTD,Kirimoto2024CTD}, which can achieve high-precision sound velocity measurement, but the observation process is time-consuming and the observation range is limited, making it difficult to achieve real-time and large-scale SSP acquisition. For example, measuring an SSP over 3000 meters depth through CTD requires at least 2 hours, during which the vessel must remain stationary. SSP inversion research provides an effective way for rapid estimation of regional sound velocity distribution, mainly including matching field processing (MFP) \cite{Tolstoy1991MFP}, compressed sensing (CS) \cite{Choo2018CS,Bianco2017CS}, and machine learning \cite{Huang2021SSPInversion,Huang2023Meta} methods. 

In 1979, Munk and Wunsch \cite{MUNK1979Tomography} first proposed the concept of ocean acoustic tomography, establishing a novel methodology for reconstructing regional sound velocity distributions through the analysis of acoustic field measurements. This approach exhibits shorter response time compared to direct measurement methods. To invert the SSP of the entire water depth, Tolstoy \cite{Tolstoy1991MFP} applies MFP to this field and combines heuristic algorithms to improve the efficiency of solving optimal solutions. Taroudakis et al. \cite{Taroudakis1996MFP} collaborated MFP with genetic algorithms (MFP-GA) for SSP inversion to improve the accuracy and efficiency. Yu et al. \cite{Yu2010MFP} validated the feasibility of the MFP-GA method in shallow waters. However, the MFP technique faces significant computational challenges, including high algorithmic complexity and prolonged processing duration, when searching the optimal matching coefficient between simulated and measured acoustic field data. To expedite the inversion process, Bianco et al. \cite{Bianco2017CS} and Choo et al. \cite{Choo2018CS} respectively proposed SSP inversion methods based on CS theory, using different types of matrices to establish the mapping relationship from sound field distribution to sound velocity distribution. But due to the introduction of linear simplification in the mapping relationship, the accuracy performance is sacrificed. 

In the past decade, the advent of artificial intelligence technology has overcome some of the limitations of traditional models and found extensive applications in marine science \cite{Bianco2019ML,Reichstein2019DLEarth}. Huang et al. \cite{Huang2018AI} proposed an SSP inversion strategy that combines artificial neural networks with ray theory, which saves time in the application phase through offline training. To reduce noise interference, they subsequently introduced an autoencoder structure and utilized autonomous underwater vehicles to construct three-dimensional sound velocity distributions \cite{Huang2021SSPInversion}. Although these neural network models exhibit better accuracy and execution efficiency compared to MFP and CS, they still rely on acoustic field measurement data, resulting in limited coverage and expensive equipment costs. 

In order to eliminate the need for underwater field data measurement and further improve the SSP estimation efficiency, the construction method of sound velocity field using remote sensing data and the prediction method of sound velocity field using historical data have become the latest research hotspots. Jain et al. \cite{Jain2006SSP} successfully demonstrated the feasibility of estimating SSPs using satellite sea surface parameters in combination with artificial neural network technology.To improve the accuracy, Li et al. \cite{Li2021SOM} proposed a self-organizing mapping (SOM) deep learning model for SSP construction that combines remote sensing sea surface temperature and sea level anomalies with empirical orthogonal function (EOF) decomposition coefficients of historical SSPs. Ou et al. \cite{Ou2022SSP} developed an end-to-end tree boosting model, Yu et al. \cite{Yu2020RBF} proposed a radial basis function neural network model, and Liu et al. \cite{LIU2023EOF} proposed a single EOF regression method that all leverage remote sensing data for SSP construction. These methods do not require underwater field data measurement, thus significantly improving the real-time performance of SSP inversion. Nevertheless, remote sensing data cannot reflect the varieties in deep sound velocity, leading to significant deviations in sound velocity estimation \cite{Kim2023RS}. 

Considering the seasonal changes in temperature and the periodic motion of ocean currents, the distribution of underwater sound velocity exhibits a certain periodic variation pattern. Therefore, some scholars have fitted the sound velocity changes from the perspective of time series prediction. Piao et al. \cite{PIAO2023IntWave} introduced long short-term memory (LSTM) neural networks for high-precision prediction of SSPs under the influence of solitary internal waves. Lu et al. \cite{Lu2024LSTM} further developed a hierarchical LSTM (H-LSTM) model, establishing separate forecasting models for each depth layer. However, LSTM focuses on the short-term variation patterns of data and is not good at capturing the long-term variation patterns of data. Moreover, as the prediction time range increases, the estimation accuracy of sound velocity distribution will significantly decrease. Recently, the proposal of Transformer model demonstrates significant advantages in time series prediction tasks \cite{Vaswani2017ATT}. The key lies in its introduction of the self-attention mechanism, which allows the model to more effectively capture long-term dependencies, thus overcoming the limitations imposed by sequence length. This is particularly critical for time series modeling, as future values may be influenced by points far back in the past. However, compared to LSTM, Transformer has a more complex network structure and higher computational complexity. To achieve accurate and real-time long-term prediction of ocean SSP without on-site data measurement, while controlling network complexity, we propose a semi-transformer neural network (STNet) model for sound velocity profile prediction in this paper. The STNet utilizes efficient attention mechanisms to comprehensively capture long-range dependencies in historical sound velocity time series data, accurately estimating the sound velocity distribution at any past time and predicting future trends in sound velocity distribution over the long term.  The main contributions of this paper are as follows:

\begin{itemize}
	\item{To achieve accurate and real-time long-term prediction of ocean SSP without on-site data measurement, we propose the STNet model for SSP prediction, which overcomes the prolonged training time associated with complex encoder-decoder structures in traditional Transformers.}
	\item{To improve execution efficiency, we proposed a parallel processing strategy during the training process of the STNet model. Time encoding and position encoding are sequentially applied to the sound velocity data to form a spatiotemporal distribution data matrix. Then the attention mechanism is used to capture the inherent dependency relationship between the temporal dynamics and spatial distribution of the data.}
	\item{To fully evaluate the effectiveness of STNet, we tested the model using historical long-period Argo observation data and short-period experimental data measured from the South China Sea in April 2023. The experimental results indicate that STNet exhibits superior performance in predicting both long and short period sound velocity distributions.}
\end{itemize}

The remainder of this paper is organized as follows. Section 2 first proposes the STNet-based structure for SSP estimation, then provides detailed working principles of important modules. Section 3 analyzes and discusses the experimental results, thoroughly evaluating the feasibility and effectiveness of STNet. Finally, conclusions are given in Section 4.

\begin{figure*}[!htbp]
	\centering
	\includegraphics[width=\linewidth]{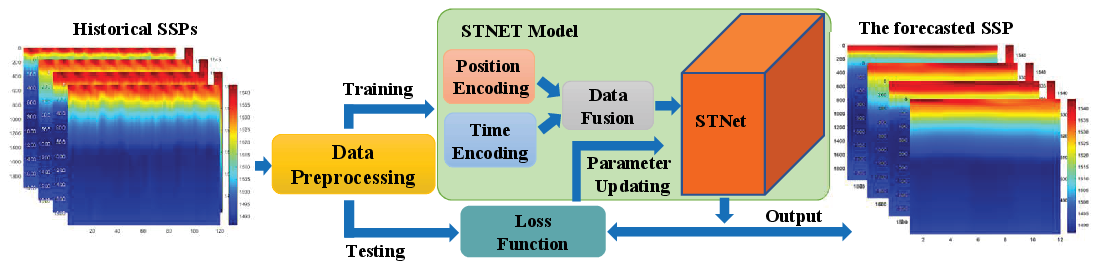}
	\caption{Framework for SSP Prediction by STNet.}
	\label{fig01}
\end{figure*}

\section{Methodology}
\subsection{Overall Framework for SSP Prediction}
In order to achieve rapid and accurate estimation or prediction of sound velocity distribution without on-site data measurement, we propose an STnet model from the perspective of time series fitting, which uses historical sound velocity to fit the sound velocity variation trend in a given area. The overall framework of SSP prediction based on STNet is given in Fig.~\ref{fig01}. 

For a specific region, historical SSP data are first sorted according to the order of sampling time, and then undergoes some preprocessing before it can be used for model training. Specifically, the data is divided into a training set and a testing set with a ratio of $8:2$. The STNet model uses parallel processing of the entire depth's SSPs to improve the efficiency of sound velocity estimation. Therefore, the data needs to be time and position encoded before being fed into the model, so that the model could better extract the correlation of sound velocity between different depth and time spans. Then, the encoded data is fused to form a data format that is easy for the model to read. In the following content, we will provide a detailed introduction to each module.

\subsection{Data and Preprocessing}
\subsubsection{Data source}
For evaluating the performance of STNet, two datasets were adopted to test the model, which were the global GDCSM\_Argo dataset \cite{Xie2019Argo} and the 2023 South China Sea experimental measured SSP (SCS-SSP) data. Benefiting from the implementation of the Argo project, a large amount of historical SSPs has been accumulated over the past decades. The GDCSM\_Argo dataset used in this paper came from the China Argo Real-time Data Center \cite{Xie2019Argo}, in which SSPs from four typical ocean regions: the South China Sea, the Atlantic Ocean, the Pacific Ocean, and the Indian Ocean, were selected as test objects. The spatial resolution of the GDCSM\_Argo dataset is $1^\circ \times 1^\circ$, the temporal resolution is one month (average SSP within a month), and the depth range is 0-1975 meters (non-uniform 58 layer sampling).

\begin{figure}[!htbp]
	\centering
	\includegraphics[width=0.8\linewidth]{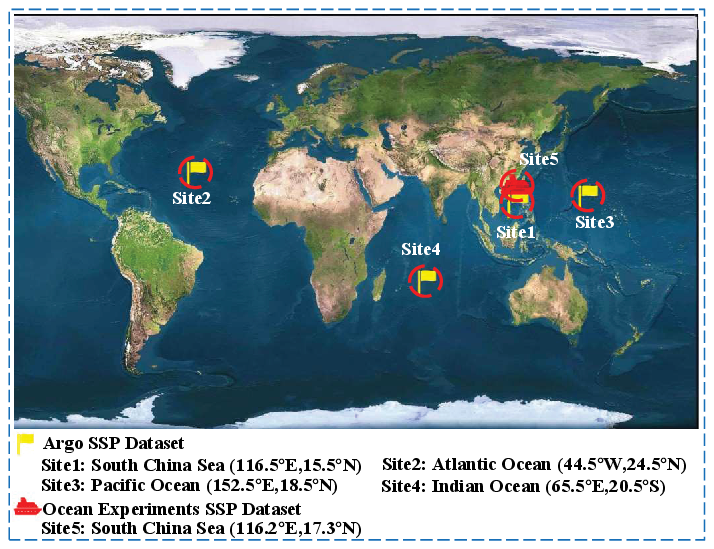}
	\caption{Locations of research area.}
	\label{fig02}
\end{figure}

To further evaluate the performance of the model, a sea trial at $17.3^\circ$N, $116.2^\circ$E in the South China Sea in April 2023 was conducted, which covered a 10 km$\times$10 km area with depth exceeding 3500 meters. The data collection system was a vessel equipped with a CTD and several expendable CTDs (XCTDs). During the experimental period, totally 14 SSPs were sampled with the average time interval as 2 hours, among which 5 SSPs were collected by CTD and 9 SSPs were collected by XCTDs. The reason for using XCTD is to save time caused by CTD measurement, but XCTD can only cover a depth range of 2000 meters. Before application, the data measured by XCTD needs to be extended to a depth of 3500 meters, which can be achieved through the method in \cite{Huang2023SSPExt}. The spatial positions of data observations are illustrated in Fig.~\ref{fig02}, with detailed information about the relevant data provided in Table 1.

\begin{table*}[!htbp]
	\caption{\textbf{Data information.}}
	\centering
	\resizebox{\textwidth}{!}{
		\begin{tabular}{cccccc}
			\toprule
			\multicolumn{6}{c}{\textbf{GDCSM\_Argo Data}} \\
			\midrule
			Area & Time dimension & Temporal resolution & Number of SSPs & Depth & Layers \\
			\midrule
			South China Sea ($116.5^\circ$E, $15.5^\circ$N) & \multirow{4}{*}{2013-2022 (120 months)} & \multirow{4}{*}{one month} & \multirow{4}{*}{120} & \multirow{4}{*}{0-1975 meters} & \multirow{4}{*}{unequal interval (58 layers)} \\
			Atlantic Ocean ($44.5^\circ$W, $24.5^\circ$N) &  &  &  &  & \\
			Pacific Ocean ($152.5^\circ$E, $18.5^\circ$N) &  &  &  &  & \\
			Indian Ocean ($65.5^\circ$E, $20.5^\circ$S) &  &  &  &  &  \\
			\midrule
			\multicolumn{6}{c}{\textbf{SCS-SSP Data}} \\
			\midrule
			South China Sea ($116.2^\circ$E, $17.3^\circ$N) & April 12-14, 2023 & Around 2 hours & 14 & 0-3500 meters & equal interval (36 layers) \\
			\bottomrule
	\end{tabular}}
	\label{table1}
\end{table*}

\subsubsection{Data resampling}
To improve the execution efficiency, the parallel computing concept is applied to STNet. However, setting an excessively high depth resolution for the sound velocity profile will lead to increased computational demands due to the larger data volume, without enhancing the model's accuracy. Therefore, we resampled the two types of data mentioned above.

In a given research area, an SSP is represented as a vector of sound velocity values across the vertical spatial distribution, such as $\mathbf{S_m} = [ s_{m,d_1},s_{m,d_2},...,s_{m,d_D} ]^T$, where $m$ represents the $i$th SSP sample, and $d_D$ represents the depth value of the $D$th depth layer. Assuming there are totally $M$ SSP samples in this region, after sorting these sound velocity data in chronological order, the dataset can be represented as follows: 
\begin{linenomath*}
\begin{equation}
	\mathcal{S}=\{\mathbf{S_1},\mathbf{S_2},...,\mathbf{S_m}\}, m=1,2,...,M. \label{eq1}
\end{equation}
\end{linenomath*}

For the GDCSM\_Argo dataset, SSPs were stratified into layers as follows: 0-10 m (5 m intervals), 10-180 m (10 m intervals), 180-460 m (20 m intervals), 500-1250 m (50 m intervals), 1300-1900 m (100 m intervals), and depths beyond 1900 m as a single layer. This preprocessing method resamples the full depth profile at different intervals, which helps to effectively reduce data dimensionality while preserving the typical structural features of the ocean SSPs. The Argo SSP data are divided into 58 layers, with the stratified data matrix organized chronologically as follows:
\begin{linenomath*}
\begin{equation}
	\mathcal{S}^{ag,rg}=\left[\begin{array}{cccc}
		s_{1,d_1}^{ag,rg} & s_{2,d_1}^{ag,rg} & \cdots & s_{120,d_1}^{ag,rg} \\
		s_{1,d_2}^{ag,rg} & s_{2,d_2}^{ag,rg} & \cdots & s_{120,d_2}^{ag,rg} \\
		& \vdots & \ddots & \vdots \\
		s_{1,d_{58}}^{ag,rg} & s_{2,d_{58}}^{ag,rg} & \cdots & s_{120,d_{58}}^{ag,rg}
	\end{array}\right],\label{eq2}
\end{equation}
\end{linenomath*}
where $rg=\text{SCS, AtO, PaO, InO}$ represents the four distinct study regions (SCS = South China Sea, AtO = Atlantic Ocean, PaO = Pacific Ocean, and InO = Indian Ocean), and $s_{120,d_{58}}^{ag,rg}$ is the sound speed value at the 58th layer of the last monthly average SSP in the ten-year GDCSM\_Argo dataset.

The ocean experiment in the South China Sea in 2023 only lasted for a few days, so the data obtained are relatively stable and covers a large depth range up to 3500 meters, weakening the influence of shallow water sound velocity changes on the overall distribution pattern. Therefore, the sea trial data from 0 to 3500 meters were stratified at equal intervals, with a total of 36 layers formed every 100 meters, excluding depths above 3500 meters. The stratified data matrix, organized chronologically, is represented as follows:
\begin{linenomath*}
\begin{equation}
	\mathcal{S}^{oe}=\left[\begin{array}{cccc}
		s_{1,d_1}^{oe} & s_{2,d_1}^{oe} & \cdots & s_{120,d_1}^{oe} \\
		s_{1,d_2}^{oe} & s_{2,d_2}^{oe} & \cdots & s_{120,d_2}^{oe} \\
		& \vdots & \ddots & \vdots \\
		s_{1,d_{36}}^{oe} & s_{2,d_{36}}^{oe} & \cdots & s_{120,d_{36}}^{oe}
	\end{array}\right],\label{eq3}
\end{equation}
\end{linenomath*}

\subsection{STNet Model}
\begin{figure}[!htbp]
	\centering
	\includegraphics[width=\linewidth]{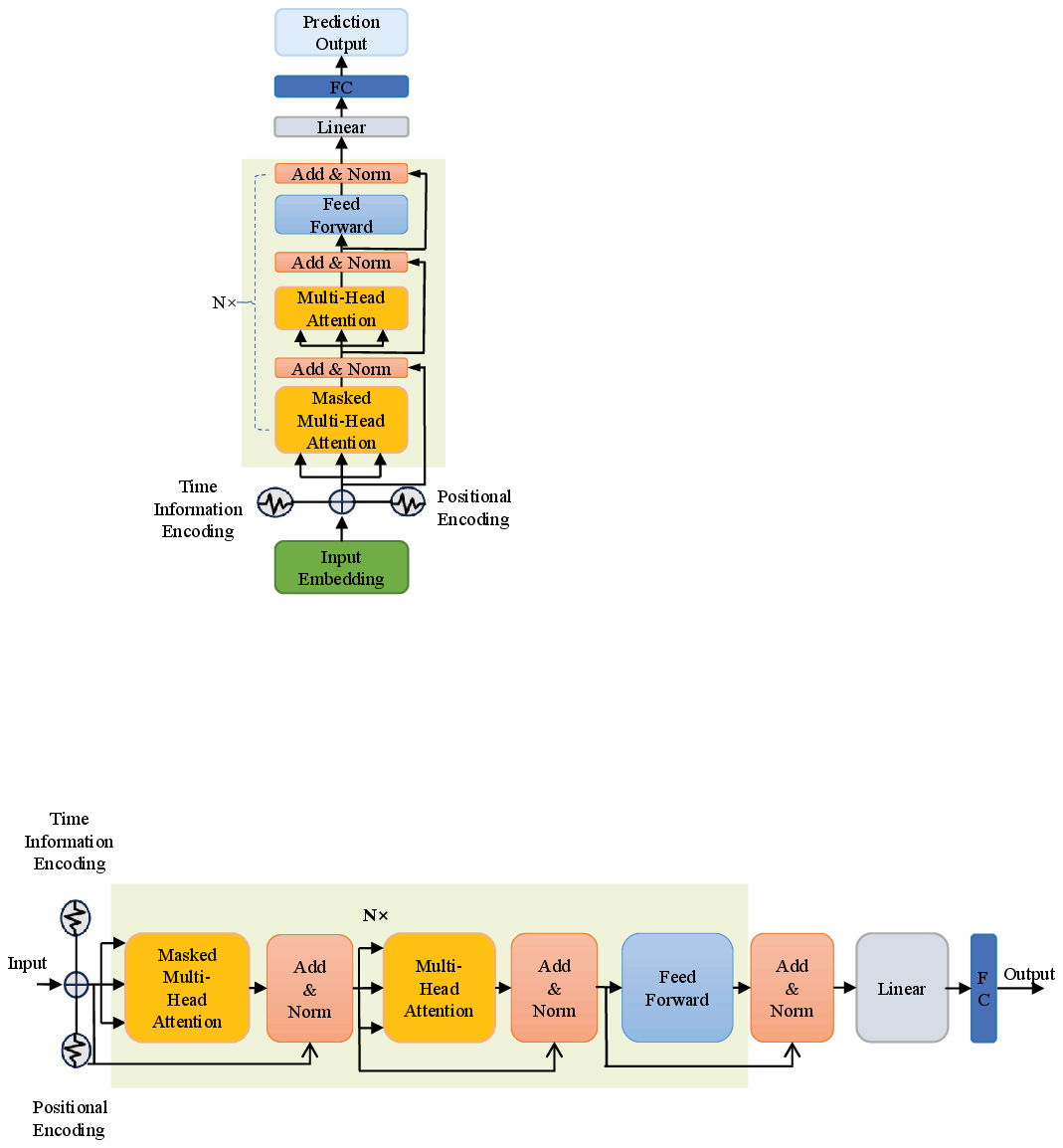}
	\caption{STNet model.}
	\label{fig03}
\end{figure}

In recent years, attention mechanisms have achieved great success in time series data processing, especially in capturing long-term dependencies. However, the famous Transformer framework composed of encoder-decoder structure appears too complex and inefficient in predicting sound velocity. To construct a simplified and effective model, we established a novel Semi-Transformer structure similar to the decoder component. Since it does not involve encoding modules, the model could significantly reduce the number of parameters and shorten training time compared to Transformer framework. 

Fig.\ref{fig03} illustrates the STNet model. To improve the feature capture capability of the model, we propose a fusion strategy of time and positional encoding. Time encoding and position encoding enable STNet to capture the trend of data changes over time while also fully considering spatial distribution correlations. Subsequently, the data undergoes 4 attention channels each with an 8-head attention module and a feed-forward neural network (FNN) layer with 128 neurons for feature extraction. Finally, a mapping relationship between features and sound velocity distribution is established through a fully connected layer. In the following subsection, we will provide a detailed introduction to each module.

\subsubsection{Time encoding}
Given that the distribution of sound velocity is significantly affected by temperature, there are significant differences in sound velocity data in different seasons, especially in shallow waters, and the same location exhibits strong similarity in the same month of different years. Temporal information encoding aims to embed specific time-related information into SSP data, similar to categorizing data by units of time such as year, month, and day. Taking Argo data as an example, the data are monthly average SSPs, so the interval for temporal encoding is month. Suppose the historical sound speed dataset contains monthly average data for N years, these data are arranged in chronological order and encoded time information starting from index 0. The time encoding is implemented as:
\begin{linenomath*}
\begin{equation}
	TE_{j}  = \frac{(j\%12)\times 2}{11}-1,
	\label{eq4}
\end{equation}
\end{linenomath*}
where $TE$ is the temporal encoding value, $j$ is the sequential index of the sound speed data with $j=0,1,...,12N-1$, and $\%$ indicates the modulo operation. Temporal encoding maps the time information within the range [-1,1]. By embedding time information, the model can distinguish data from the same month in different years.

\subsubsection{Positional encoding}
To improve the efficiency of model operation, we propose a parallel processing strategy. In this way, time series data of sound velocity at different depths can be processed simultaneously. Due to the differences in the time series patterns of sound velocity in different depth layers and the typical vertical distribution of sound velocity profile data, it is particularly important to encode the position of the sound velocity time series, which can help the model understand the sequential relationship between time series data in different depth layers. The positional encoding can be achieved by:
\begin{linenomath*}
\begin{equation}
	\begin{cases}
		PE_{p,2i} & =  \sin \left(\frac{p}{10000^{\frac{2i}{F}}}\right)\\
		PE_{p,2i+1} & = \cos \left(\frac{p}{10000^{\frac{2i}{F}}}\right)
	\end{cases},\label{eq5}
\end{equation}
\end{linenomath*}
where $p$ represents the position in the sequence, $i$ is the dimension index in the positional encoding vector, and $F$ is the dimension of the input vector.

\subsubsection{Self-attention mechanism}
The self-attention mechanism is an innovative approach that excels in capturing long-range dependencies, making it suitable for the task of predicting ocean sound speed distribution. Let $\mathcal{S}$ is the input sound speed time series variables after position and temporal encoding. By linearly transforming $\mathcal{S}$ through three weight matrices $\mathcal{W}_Q$, $\mathcal{W}_K$ and $\mathcal{W}_V$, the Query matrix $\mathcal{Q}$, the Key matrix $\mathcal{K}$ and the Value matrix $\mathcal{V}$ for calculating attention values can be obtained respectively:
\begin{linenomath*}
\begin{equation}
	\begin{cases}
		\mathcal{Q} =  \mathcal{S}\mathcal{W}_Q\\
		\mathcal{K} =  \mathcal{S}\mathcal{W}_K\\
		\mathcal{V} =  \mathcal{S}\mathcal{W}_V
	\end{cases}.\label{eq6}
\end{equation}
\end{linenomath*}
Once $\mathcal{Q}$, $\mathcal{K}$ and $\mathcal{V}$ are obtained, the attention scores can be calculated through:
\begin{linenomath*}
\begin{equation}
	Att(\mathcal{Q}, \mathcal{K}, \mathcal{V}) = softmax(\frac{\mathcal{Q}\mathcal{K}^T}{\sqrt{F_K}})\mathcal{V},\label{eq7}
\end{equation}
\end{linenomath*}
where $F_K$ is the dimension of $\mathcal{K}$.

\begin{figure}[!htbp]
	\centering
	\includegraphics[width=0.5\linewidth]{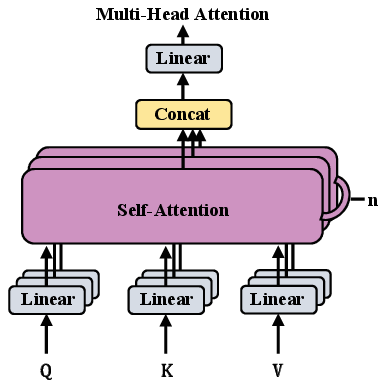}
	\caption{Multi-head attention module.}
	\label{fig04}
\end{figure}

\paragraph{Multi-head attention mechanism}
To simultaneously focus on different subspace information of the input sequence, the implementation of the attention mechanism actually adopts a multi-head attention structure as shown in Fig.\ref{fig04}.

Fig.\ref{fig04} introduces multiple independently learning attention heads in parallel in one attention channel, enabling the model to analyze sound velocity data from different subspaces and comprehensively capture key features in the data. Through the weighted integration of multi-subspace feature representations, the model enhances its capability to capture intrinsic correlations within acoustic velocity time series, consequently improving the predictive accuracy of sound velocity distribution patterns. That is to say, multiple sets of $\mathcal{Q}$, $\mathcal{K}$ and $\mathcal{V}$ are obtained with the help of different sets of $\mathcal{W}_Q$, $\mathcal{W}_K$ and $\mathcal{W}_V$ matrices. Then, the calculated attention values of each attention head are combined through weighted combination and linear transformation. In this way, the model can focus on the importance of different positions in the input sound velocity sequence, while capturing local and global dependencies.

Suppose there are $U$ attention heads in each attention channel. According to \eqref{eq7}, the attention value of the $u$th head will be:
\begin{linenomath*}
\begin{equation}
	\mathcal{H}_u = Att(\mathcal{Q}_u, \mathcal{K}_u, \mathcal{V}_u),\label{eq8}
\end{equation}
\end{linenomath*}
where $\mathcal{Q}_u$, $\mathcal{K}_u$ and $\mathcal{V}_u$ are the Query, Key, and Value matrices of the $u$th attention head. The final output feature of the multi-head attention channel is:
\begin{linenomath*}
\begin{equation}
	\mathcal{MH}(\mathcal{Q}, \mathcal{K}, \mathcal{V}) = Concat(\mathcal{H}_1,\mathcal{H}_2,...,\mathcal{H}_U)\mathcal{W}_0,\label{eq9}
\end{equation}
\end{linenomath*}
where $Concat$ is a linking function, and $\mathcal{W}_0$ is a linear transformation matrix.

\paragraph{Masked Multi-head attention mechanism}
The calculation principles of Masked Multi-Head Attention and Multi-Head Attention are consistent, with the only difference being the addition of a mask code. When processing time series, the output of the current time $t$ should only depend on the output before time $t$, and not on the output after $t$. The mask code ensures that future time information will not be accessed by the model during training.

\subsubsection{Feed-forward Neural Network Layer}
Besides the attention layers, the STNet stack module includes a FNN layer. Let $\mathcal{Y}_{MH}$ represents the output of the attention layer. The result $\mathcal{Y}_{F}$ after passing through the FFN can be expressed as:
\begin{linenomath*}
\begin{equation}
	\mathcal{Y}_{F} = max(0,\mathcal{Y}_{MH}\mathcal{W}_1+\mathbf{b_1})\mathcal{W}_2+\mathbf{b_2},\label{eq10}
\end{equation}
\end{linenomath*}
where $\mathcal{W}_1$ and $\mathcal{W}_2$ are linear transformation matrices, and $\mathbf{b_1}$ and $\mathbf{b_2}$ are biases. It is evident that $\mathcal{Y}_{F}$ undergoes two linear transformations and one ReLU activation function.

After passing through the attention layer and FFN layer, the output features are not directly transmitted to the next layer, but undergo residual processing and data normalization. The purpose of this design is to prevent degradation during model training, while accelerating training speed and improving training stability.

\subsubsection{Model Parameter Updating}
For a task region, a dedicated model will be constructed according to Fig.\ref{fig03}, and the training process will be completed offline in advance so as to save the time cost of applying the model in predicting SSPs. To improve training efficiency and accelerate model convergence, we adopts batch gradient descent and adaptive learning rate adjustment strategies.

Suppose that the SSP estimated by the model is $\mathbf{\hat{S}} = [\hat{s}_{t+1,d_1},\hat{s}_{t+1,d_2},...,\hat{s}_{t+1,d_z}]^T$, $z=1,2,...,Z$, where $t+1$ represents the next timestamp, and there are total $Z$ depth layers, the parameters of the model will be updated by back propagation algorithm according to loss function:
\begin{linenomath*}
\begin{equation}
	L_{rmse} = \sqrt{\frac{\sum_{z=1}^{Z}\left(\hat{s}_{t+1,d_z}-s_{t+1,d_z}\right)^{2}}{Z}}, z=1,2, \ldots, Z,\label{eq11}
\end{equation}
\end{linenomath*}
where $s_{t+1,d_z}$ and $\hat{s}_{t+1,d_z}$ respectively represent the actual and estimated sound velocity values at the $z$th depth layer. 

\section{Results and Discussions}
\subsection{Parameter Settings and Baselines}
The data processing and model validation in this article were completed in MATLAB R2023b with model parameter settings listed in Table~\ref{table2}, which were obtained through multiple experimental attempts. To comprehensively evaluate the reliability of the STNet model, H-LSTM \cite{Lu2024LSTM}, multi-layer perceptron (MLP) \cite{Yu2020RBF}, and polynomial fitting (PF) \cite{Liu2019PF} are selected as baselines for comparisons in this paper. Besides the root mean square error (RMSE) of \eqref{eq11} that $L_{rmse}$, mean square error (MSE) $L_{mse}$ and mean absolute error (MAE) $L_{mae}$ are also adopted as evaluation metrics:
\begin{linenomath*}
\begin{equation}
	L_{mse} = \frac{\sum_{z=1}^{Z}\left(\hat{s}_{t+1,d_z}-s_{t+1,d_z}\right)^{2}}{Z}, z=1,2, \ldots, Z,\label{eq12}
\end{equation}
\end{linenomath*}
\begin{linenomath*}
\begin{equation}
	L_{mae} = \frac{\sum_{z=1}^{Z}\left|\hat{s}_{t+1,d_z}-s_{t+1,d_z}\right|}{Z}, z=1,2, \ldots, Z.\label{eq13}
\end{equation}
\end{linenomath*}

\begin{table}[!htbp]
	\caption{\textbf{Model Parameter Settings.}}
	\centering
	\begin{tabular}{cc}
		\toprule
		\textbf{Parameter} & \textbf{Setting}\\
		\midrule
		Dimension of sequence input layer & 58/36 \\
		\midrule
		Number of heads & 8 \\
		\midrule
		Number of attention channels & 4 \\
		\midrule
		Neurons of FNN layer & 128 \\
		\midrule
		Dropout rate & 0.15 \\
		\midrule
		Max epoch & 300 \\
		\midrule
		Batch size & 32 \\
		\midrule
		Optimizer & Adam \\
		\midrule
		Initial learning rate & 0.001 \\
		\bottomrule
	\end{tabular}
	\label{table2}
\end{table}

\subsection{Influence of Training Time Stepping}
To optimize the selection of training sequence steps, we conducted multiple experiments taking the Pacific region as an example. Table~\ref{table3} provides the prediction accuracy of the model for the monthly sound velocity distribution in the next year under different training steps. The results are the average of 50 attempts. The results show that in most months, when the training time step is 1, the prediction performance is better than the others, with an average prediction error of 0.5811 m/s. Therefore, the model training time step would be 1 in subsequent experiments. An interesting phenomenon is that the predicted value for the 12th month in the future is significantly better than that for the 11th month, which may be influenced by the annual cycle variation of sound velocity.

\subsection{Influence of Training Data Length}
To test the influence of different historical SSP data time spans on the prediction accuracy performance of the model, we took the Argo dataset in the Pacific ocean as an example and selected data from the first 1, 3, 5, 7, and 9 years before the prediction task for model training. The average prediction error of 50 attempts is given in Table~\ref{table4}. The results indicate that as the training data time span increases, the predictive performance of the model gradually improves. When there are too few learning samples, the model struggles to capture the trend of data changes, leading to an increase in prediction error RMSE. Specifically, at least 5 or more complete cycles of SSP data before the forecast task need to be provided for model learning in order to ensure relatively accurate predictions for future SSP.

\begin{table*}[!htbp]
	\caption{\textbf{Prediction errors of SSPs under different training steps.}}
	\centering
	\resizebox{\linewidth}{!}{
		\begin{tabular}{cccccccccccccc}
			\toprule
			& \multicolumn{12}{c}{\textbf{RMSE for different months (m/s)}} & \\
			\cline{2-13}
			\textbf{Training Time Step$\setminus$ Month} & 1 & 2 & 3 & 4 & 5 & 6 & 7 & 8 & 9 & 10 & 11 & 12 & \textbf{Average RMSE (m/s)} \\
			\midrule
			1 time step & 0.637 & 0.563 & 0.751 & 0.405 & 0.445 & 0.983 & 0.444 & 0.675 & 0.422 & 0.813 & 0.529 & 0.309 & \textbf{0.581}\\
			\midrule
			2 time steps & 0.602 & 0.546 & 0.808 & 0.534 & 0.784 & 0.965 & 0.904 & 0.950 & 0.350 & 0.923 & 1.307 & 0.495 & \textbf{0.763}\\
			\midrule
			4 time steps & 0.776 & 0.478 & 0.468 & 0.653 & 1.270 & 1.338 & 0.485 & 0.930 & 0.536 & 0.662 & 0.998 & 0.577 & \textbf{0.764}\\
			\midrule
			6 time steps & 0.705 & 0.706 & 0.800 & 0.851 & 0.557 & 0.846 & 0.784 & 0.827 & 0.752 & 1.040 & 1.132 & 0.470 & \textbf{0.789}\\
			\midrule
			10 time steps & 0.996 & 1.241 & 1.357 & 1.253 & 0.731 & 0.848 & 0.455 & 0.662 & 0.569 & 0.702 & 0.663 & 0.584 & \textbf{0.838}\\
			\bottomrule
		\end{tabular}
	}
	\label{table3}
\end{table*}
\begin{table*}[!htbp]
	\caption{\textbf{Prediction errors of SSPs under different training data lengths.}}
	\centering
	\resizebox{\linewidth}{!}{
		\begin{tabular}{cccccccccccccc}
			\toprule
			& \multicolumn{12}{c}{\textbf{RMSE for different months (m/s)}} & \\
			\cline{2-13}
			\textbf{Training Data Length$\setminus$ Month} & 1 & 2 & 3 & 4 & 5 & 6 & 7 & 8 & 9 & 10 & 11 & 12 & \textbf{Average RMSE (m/s)} \\
			\midrule
			1 year & 1.647 & 0.529 & 0.805 & 0.795 & 0.858 & 0.813 & 1.175 & 1.041 & 0.880 & 0.949 & 0.857 & 0.816 & \textbf{0.930}\\
			\midrule
			3 years & 0.521 & 0.678 & 1.320 & 0.584 & 0.980 & 1.288 & 0.781 & 0.711 & 0.779 & 0.767 & 0.739 & 0.837 & \textbf{0.832}\\
			\midrule
			5 years & 0.667 & 0.755 & 0.934 & 0.854 & 0.556 & 1.057 & 0.621 & 0.405 & 0.548 & 0.550 & 0.432 & 0.583 & \textbf{0.664}\\
			\midrule
			7 years & 0.698 & 0.537 & 0.557 & 0.689 & 0.465 & 1.023 & 0.629 & 0.601 & 0.373 & 0.453 & 0.487 & 0.549 & \textbf{0.588}\\
			\midrule
			9 years & 0.637 & 0.563 & 0.751 & 0.405 & 0.445 & 0.983 & 0.444 & 0.675 & 0.422 & 0.813 & 0.529 & 0.309 & \textbf{0.581}\\
			\bottomrule
		\end{tabular}
	}
	\label{table4}
\end{table*}

\subsection{Evaluation of Periodic Capture Capability}
\begin{figure}[!htbp]
	\centering
	\includegraphics[width=\linewidth]{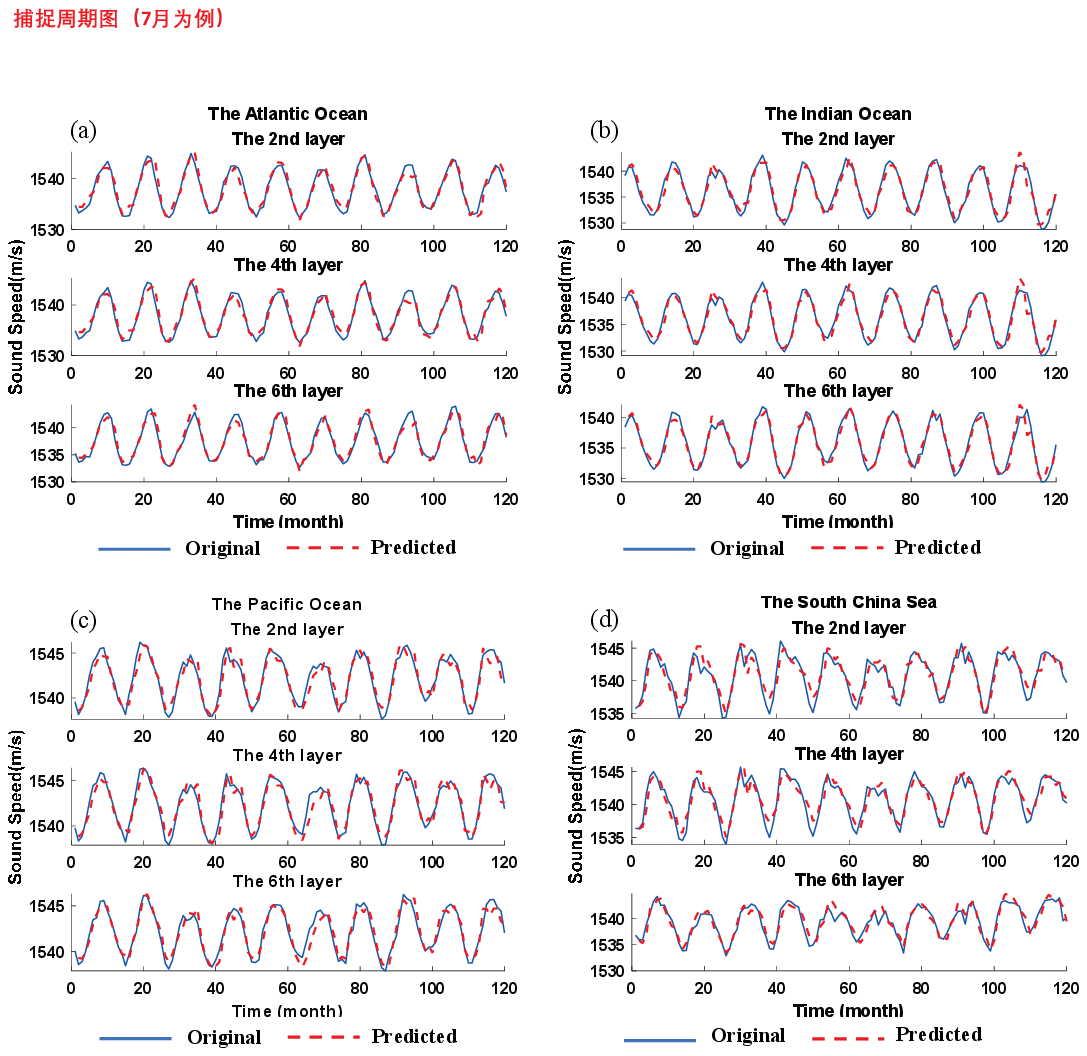}
	\caption{Periodic variation fitting results, where (a)(b)(c)(d) are the result of the Atlantic Ocean, Indian Ocean, Pacific Ocean, and South China Sea, respectively.}
	\label{fig05}
\end{figure}

To validate whether STNet can accurately capture the periodic variations in sound speed time series, we tested the model using ten years of historical data from four regions according to Table~\ref{table1}. The 2nd, 4th, and 6th layers (corresponding to depths of 5 meters, 20 meters, and 40 meters, respectively) were randomly selected from the 58 depth layers to demonstrate the model's training and prediction outputs as shown in Fig.\ref{fig05}. The comparison of periodic variation trends at different depth layers across the four regions demonstrates that STNet has good abilities to learn long-range dependencies with the input sequences.

\begin{figure}[!htbp]
	\centering
	\includegraphics[width=\linewidth]{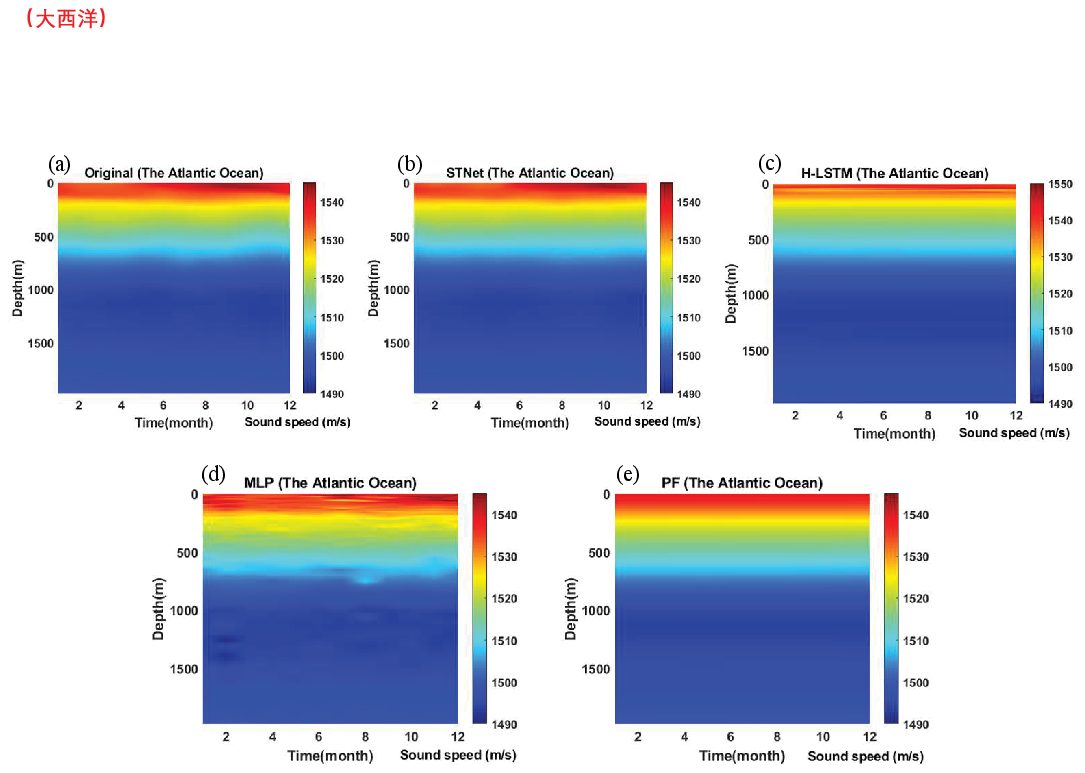}
	\caption{Accuracy performance of long-term SSP forecasting in the Atlantic Ocean, where (a) is the original sound velocity distribution, (b)(c)(d)(e) are the estimated results by STNet, H-LSTM, MLP and PF, respectively.}
	\label{fig06}
\end{figure}

\begin{figure}[!htbp]
	\centering
	\includegraphics[width=\linewidth]{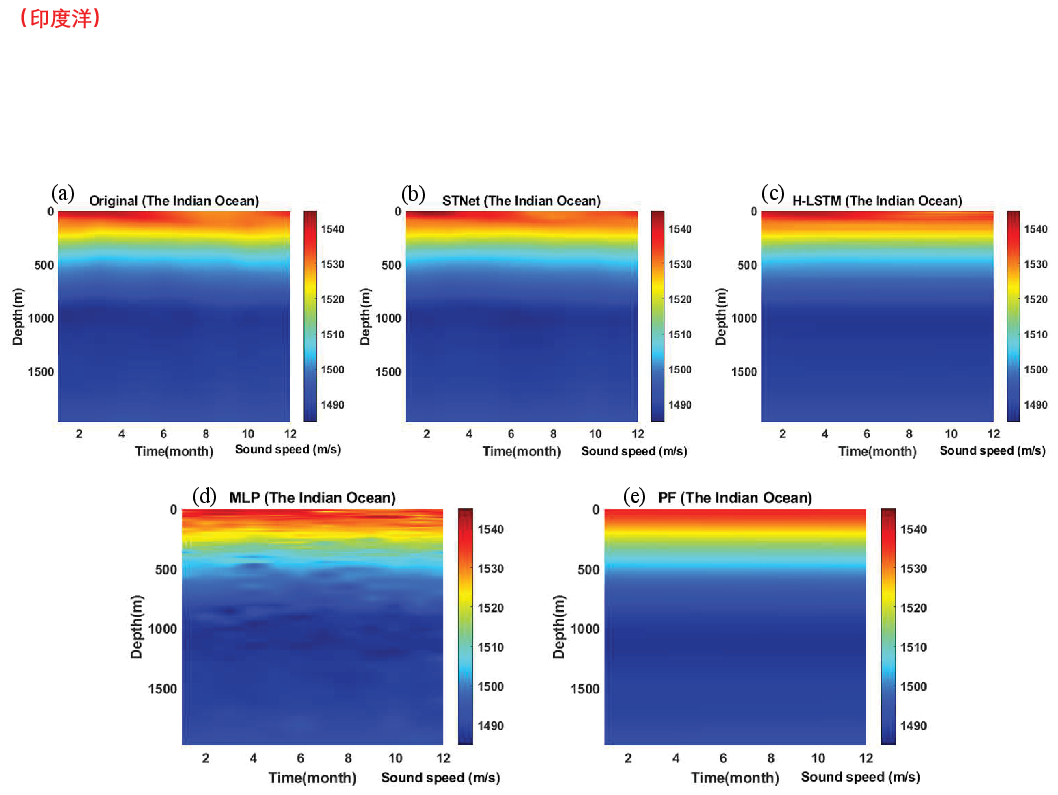}
	\caption{Accuracy performance of long-term SSP forecasting in the Indian Ocean, where (a) is the original sound velocity distribution, (b)(c)(d)(e) are the estimated results by STNet, H-LSTM, MLP and PF, respectively.}
	\label{fig07}
\end{figure}

\subsection{Long-Term Predictive Performance Evaluation}
To evaluate the accuracy performance of the model in long-term forecasting, the first 9 years of historical Argo datasets from the Atlantic Ocean and the Indian Ocean are used as learning samples for the model to predict the sound velocity distribution for the corresponding regions in the next year. The training data consisted of 58 unequally layered SSP data (formatted as [58,108]), and the primary prediction results were also divided into 58 layers (formatted as [58,12]). After interpolation with a spacing of 1 meter, the full-depth results (formatted as [1976,12]) were compared with the baseline methods in Fig.\ref{fig06} and Fig.\ref{fig07}. Comparatively speaking, the prediction results of H-LSTM and PF methods to some extent reflect the distribution of underwater SSP, but the prediction accuracy is not good enough, so these methods can be used for preliminary estimation of underwater SSPs. The prediction results of the MLP model are relatively rough and cannot accurately reflect the actual distribution of SSP in complex marine environments. In contrast the predicted results of STNet are highly consistent with actual observed data, demonstrating higher accuracy in the long-term prediction task of deep Ocean SSP. Specifically, in complex shallow sea environments (within 200 meters of sea depth), STNet can effectively capture the variation patterns of marine SSP.

\begin{figure}[!htbp]
	\centering
	\includegraphics[width=\linewidth]{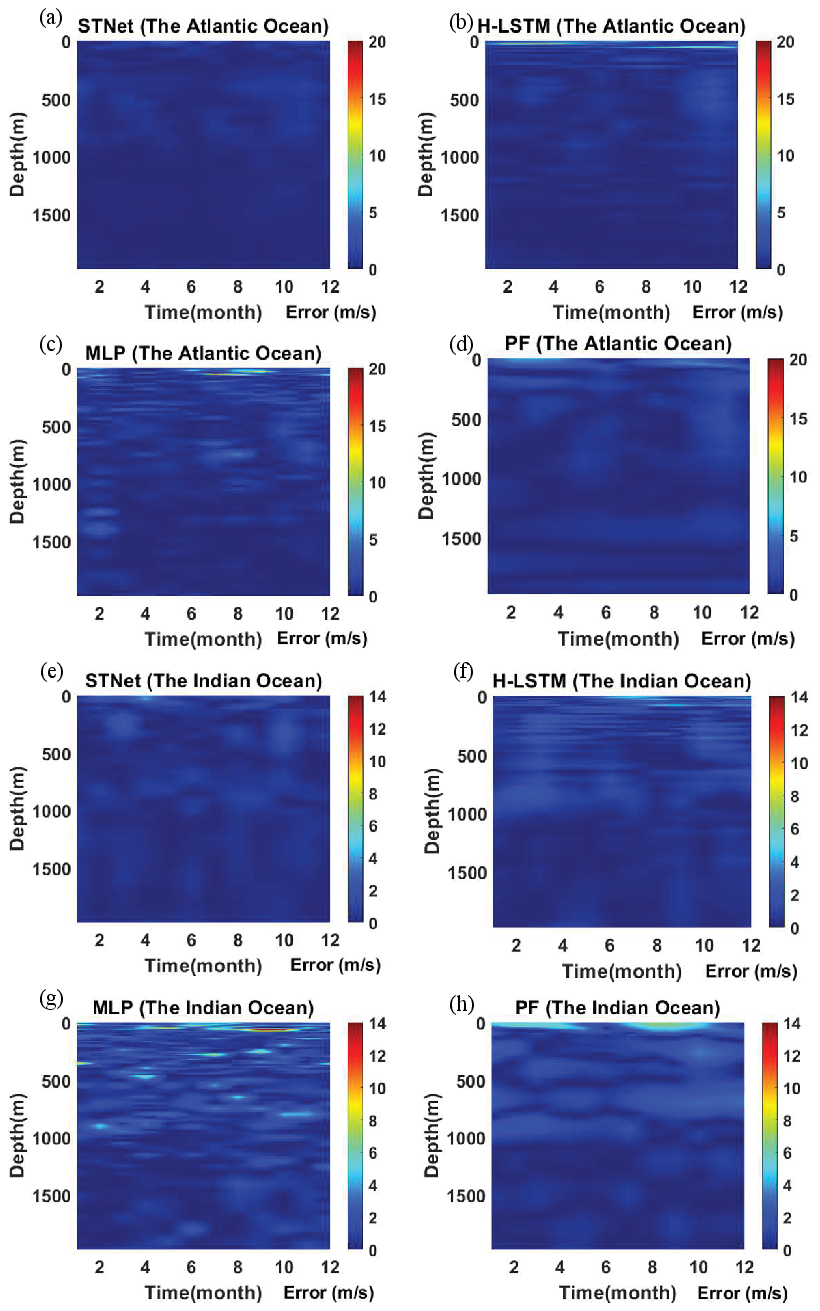}
	\caption{Absolute errors of long-term SSP prediction at different depths, where (a)(b)(c)(d) are the estimated errors by STNet, H-LSTM, MLP and PF in the Atlantic Ocean, and (e)(f)(g)(h) are the estimated errors by STNet, H-LSTM, MLP and PF in the Indian Ocean.}
	\label{fig08}
\end{figure}

To more clearly assess the prediction performance of various models, Fig.\ref{fig08} illustrates the absolute errors at different depths for the Atlantic Ocean and the Indian Ocean. In the Atlantic Ocean, the maximum absolute errors of STNet, H-LSTM, MLP and PF at full ocean depth are approximately 3 m/s, 11 m/s, 20 m/s, and 5.5 m/s, respectively. The absolute errors of these four methods at most depths are below 0.5 m/s, 2 m/s, 4 m/s, and 1.5 m/s, respectively. In the Indian Ocean, the maximum absolute errors of STNet, H-LSTM, MLP and PF at full ocean depth are approximately 4 m/s, 5.5 m/s, 12 m/s, and 7 m/s, respectively. The absolute errors at most depths are below 0.5 m/s, 1 m/s, 2 m/s, and 1 m/s, respectively. Overall, the absolute error of the STNet model in predicting sound velocity at different depths in the Atlantic and Indian Oceans is significantly lower than that of the H-LSTM, BP, and PF models, demonstrating its superiority in capturing the details of sound velocity time series.

\begin{figure}[!htbp]
	\centering
	\includegraphics[width=\linewidth]{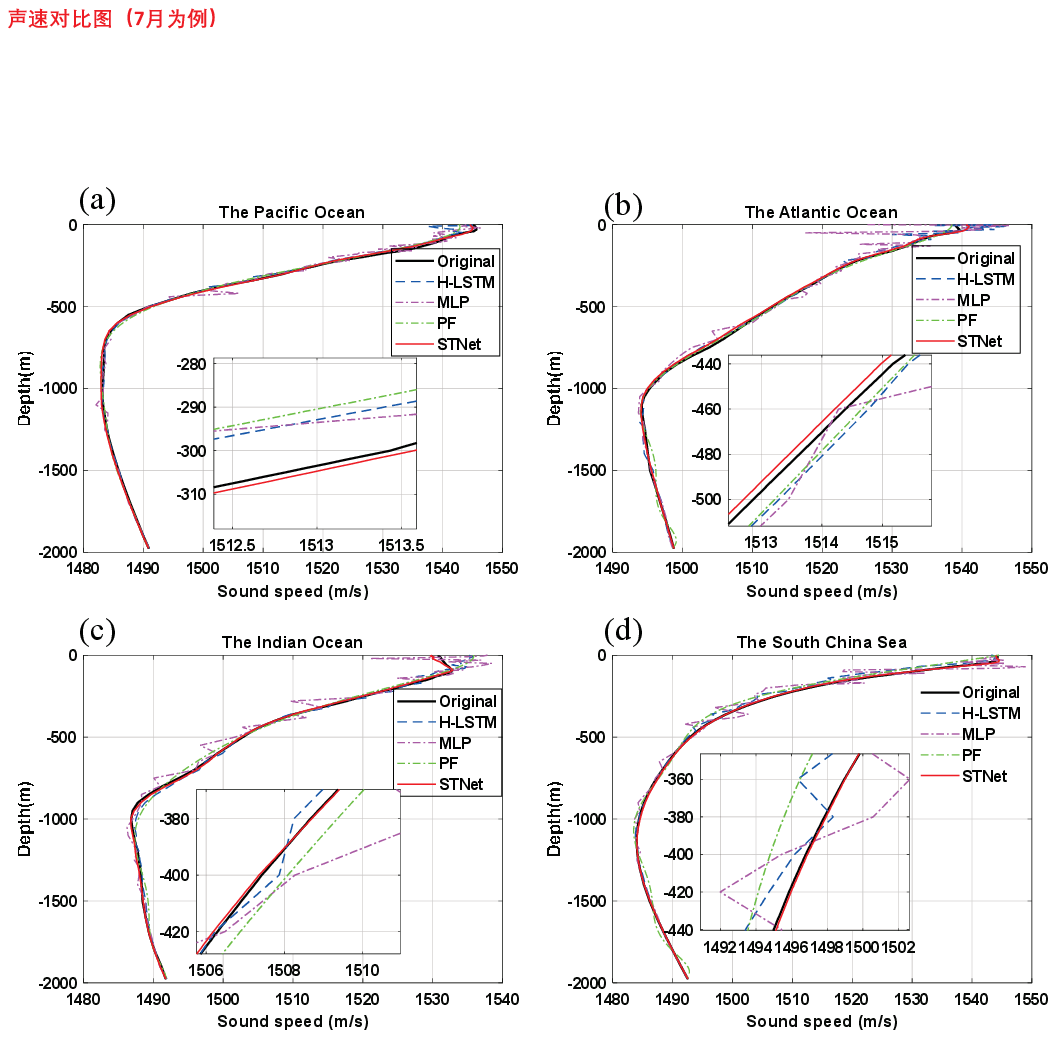}
	\caption{Comparison of predicted and actual SSPs in July, 2022 for the four ocean regions, where (a)(b)(c)(d) correspond to the Atlantic Ocean, Indian Ocean, Pacific Ocean, and South China Sea, respectively.}
	\label{fig09}
\end{figure}

To visually compare the differences between the predicted SSP and the actual SSP for various models, Fig.\ref{fig09} presents a two-dimensional comparison of predicted and actual SSPs in July, 2022 for the four ocean regions referring to Table~\ref{table1}. The results show that among the four sea areas, the SSP predicted by STNet has the best fit with the actual curve, followed by H-LSTM and PF, while MLP has the largest fitting error. The SSP predicted by the STNet model can accurately reflect the characteristic information of the actual SSP, and is more stable in long-term prediction of the sound velocity.

\begin{figure}[!htbp]
	\centering
	\includegraphics[width=\linewidth]{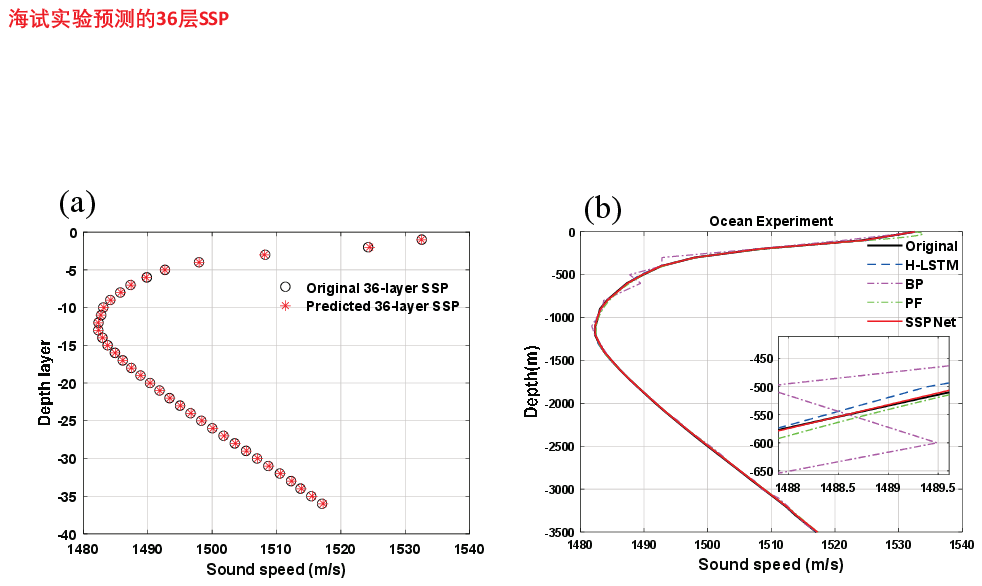}
	\caption{Comparison of predicted and actual SSPs in the South China Sea in 2023, where (a) is comparison between the SSP predicted by the STNet model with the observed SSP, and (b) is comparison among different methods after interpolating.}
	\label{fig10}
\end{figure}

\subsection{Short-Term Predictive Performance Evaluation}

To validate the short-term prediction capability of the STNet model, we conducted short-term sound velocity data collection in the South China Sea in 2023. The model was trained by the first 13 sampled SSPs to predict the sound velocity distribution for the next two hours. Fig.\ref{fig10}(a) directly provides a comparison between the predicted SSP and the label one, which has a high similarity. In Fig.\ref{fig10}(b), the SSP is interpolated and compared with the results obtained by the H-LSTM, MLP and PF methods. The results indicate that the STNet model's predictions closely match the actual SSPs across the 0-3500 meter depth range, exhibiting high similarity. The H-LSTM model also performs well, but it lags behind STNet in handling finer details. The PF model could roughly capture the global characteristics of the actual SSPs, but shows significant deviations in shallow waters (within 200 meters). The MLP model does not perform well in dealing with time series, resulting in predictions with high fluctuations, failing to accurately reflect the primary features of the actual SSPs. Table~\ref{table5} shows the corresponding full-depth prediction errors under 50 attempts. The results show that compared to H-LSTM, MLP, and PF models, STNet significantly improves SSP prediction accuracy, reaffirming the superior capability of our new model in SSP prediction.

\begin{table}[!htbp]
	\caption{Short-Term SSP prediction errors.}
	\centering
	\begin{tabular}{ccccc}
		\toprule
		Method & STNet & H-LSTM & MLP & PF \\
		\midrule
		RMSE (m/s) & \textbf{0.079} & 0.153 & 0.957 & 0.548 \\
		\bottomrule
	\end{tabular}
	\label{table5}
\end{table}

\subsection{Comparison of Execution Efficiency}
The proposed STNet employs a parallel processing strategy, enabling the simultaneous handling of multiple sound speed time series data. This approach is of great significance for improving the efficiency of SSP prediction. Table~\ref{table6} shows a detailed comparison of the average time efficiency of different methods during the model training phase under 50 attempts. For STNet, with parallel processing, it takes only 28.07 seconds for model training, which is nearly 10 times more efficient than deep learning models like H-LSTM and MLP. Even compared to the mathematically efficient PF method, the difference is minimal. This result highlights the efficiency of parallel processing strategies, indicating the enormous potential of STNet in predicting ocean sound velocity. 

\begin{table}[!htbp]
	\caption{SSP prediction errors of ocean experiment in the South China Sea.}
	\centering
	\begin{tabular}{ccccc}
		\toprule
		Method & STNet & H-LSTM & MLP & PF \\
		\midrule
		Training time (s) & 28.07 & 223.19 & 232.30 & 12.33 \\
		\bottomrule
	\end{tabular}
	\label{table6}
\end{table}

\section{Conclusion}
To achieve real-time and accurate long-term prediction of full-depth ocean SSPs, we proposes an STNet model. STNet simplifies and optimizes traditional Transformer models while cleverly incorporating time encoding modules. This design enables STNet to effectively capture long-range dependencies within historical sound speed time series data and accurately track temporal trends. To validate the feasibility and effectiveness of the STNet model, extensive long-term prediction experiments were conducted across multiple ocean regions. Experimental results indicate that STNet outperforms other state-of-the-art models in both prediction accuracy and time efficiency. Consequently, STNet provides crucial technical support for achieving precise, large-scale, and full-depth predictions of ocean sound speed profiles.
%
%

\section*{Open Research Section}
The authors acknowledge the historical SSP data support from the China Argo Real-time Data Center, (https://www.argo.org.cn/, latest access:January 10, 2025). Code can be obtained from https://github.com/WeiWilliamHuang/SemiTrans.git. The authors declare that there is no conflict of interest.

\acknowledgments
This work was supported in part by the National Key Research and Development Program of China (2024YFB3909701), in part the National Natural Science Foundation of China under Grant 42404001 and 62271459, in part by Natural Science Foundation of Shandong Province under Grant ZR2023QF128. There is no conflict of interests.

%
%

\bibliography{draft.bib}

\end{document}